\documentclass[aps,prd,twocolumn,showpacs,superscriptaddress,groupedaddress]{revtex4} 
\usepackage{graphicx}
\usepackage{dcolumn}
\usepackage{bm}

\newcommand\br{\begin{eqnarray}}
\newcommand\er{\end{eqnarray}}
\newcommand\be{\begin{equation}}
\newcommand\ee{\end{equation}}

\newcommand\bc{\begin{center}}
\newcommand\ec{\end{center}}

\newcommand{\nn}{\nonumber \\}

\newcommand\Tr{\mathop{\mathrm Tr}}                  

\newcommand{\tram}{\Tr(T^{a}U\partial_{\mu}U^{-1})}
\newcommand{\trbm}{\Tr(T^{b}U\partial_{\mu}U^{-1})}

\newcommand{\tran}{\Tr(T^{a}U\partial_{\nu}U^{-1})}

\newcommand{\trcn}{\Tr(T^{c}U\partial_{\nu}U^{-1})}

\begin{document}
\title{Confining Boundary conditions from dynamical Coupling Constants for Abelian and non Abelian symmetries}

\author
{R. Steiner \footnote{e-mail: roeexs@gmail.com}}
\address{Physics Department, Ben Gurion University of the Negev, Beer
Sheva 84105, Israel}

\author
{E. I. Guendelman \footnote{e-mail: guendel@bgu.ac.il}}
\address{Physics Department, Ben Gurion University of the Negev, Beer
Sheva 84105, Israel}
\date{\today}
\begin{abstract}
In this paper we present a model which can produce boundary confining condition on Dirac field interacting with Abelian or non Abelian gauge fields. The constraint is generated by a scalar field. This kind of model can be the foundation for bag models which can produce confinement. The present work represents among other things a generalization to the non Abelian case of our previous result where the Abelian case was studied. In the $U(1)$ case the coupling to the gauge field  contains a term of the form $g(\phi)j_\mu (A^{\mu} +\partial^{\mu}B)$ where $B$ is an auxiliary field and  $j_\mu$ is the Dirac current. The scalar field $\phi$ determines the local value of the coupling of the gauge field to the Dirac particle. The consistency of the equations determines the condition $\partial^{\mu}\phi j_\mu  = 0$
which implies that the Dirac current cannot have a component in the direction of the gradient of the scalar field. As a consequence, if  $\phi$ has a soliton behavior, like defining a bubble that connects two vacuua, we obtain that the Dirac current cannot have a flux through the wall of the bubble, defining a confinement mechanism where the fermions are kept inside those bags. In this paper we present more models in Abelian case which produce constraint on the Dirac or scalar current and also spin. Furthermore a model that give the M.I.T confinement condition for gauge fields is obtained. We generalize this procedure for the non Abelian case and we find a constraint that can be used to build a bag model. In the non Abelian case the confining boundary conditions hold at a specific surface of a domain wall.
\end{abstract}
\maketitle

\section{Introduction}
In this paper we present a model which produce boundary constraint on Abelian and non Abelian Dirac field. The constraint controlled by a real scalar field or a complex scalar field. This kind of model can be the basis a for bag model which can describe confinement. The concept is like that of our previous paper \citep{GlobalSQED} which will review here. This model presents the possibility of deriving boundary condition from the action principle in the Abelian gauge symmetry case. In that paper we added global gauge invariant term which couple to dynamical coupling constant. The coupling to the gauge field  contains a term of the form $g(\phi)j_\mu (A^{\mu} +\partial^{\mu}B)$ where $B$ is an auxiliary field and  $j_\mu$ is the Dirac current. The scalar field $\phi$ determines the local value of the coupling of the gauge field to the Dirac particle. The consistency of the equations determines the condition $\partial^{\mu}\phi j_\mu  = 0$ which implies that the Dirac current cannot have a component in the direction of the gradient of the scalar field. As a consequence, if  $\phi$ has a soliton behavior, like defining a bubble that connects two vacuua, we obtain that the Dirac current cannot have a flux through the wall of the bubble, defining a confinement mechanism where the fermions are kept inside those bags. In this paper we present more models in Abelian case which produce constraint on the Dirac or scalar current and also spin. Furthermore a model that give the M.I.T confinement condition for gauge fields is obtained. We generalize this procedure for the non Abelian case and we find a constraint that can be used to build a bag model. In the non Abelian case the confining boundary conditions hold at a specific surface of a domain wall.\\
In the past there were confinement models, the basic idea of these models was the confinement of quarks in a cavity of finite size, this cavity usually taken to be spherical symmetric. Some of the models like the MIT bag model \citep{mitbag, mitbag1} , added boundary conditions to the action which provided zero color charge and therefore a confinement mechanism.\\
Because of the non natural treatment of adding boundary condition to the action Friedberg and Lee \citep{FriedbergandLee, FriedbergandLee1} have constructed a non-topological soliton realizetion of the bag model in the presence of a non-uniform dielectric medium.\\
The dielectric constant is a function of a scalar field $ \sigma $ and the confining phase is identified as the region where the dielectric constant becomes zero. This produces in fact an infinite effective coupling constant in the confinement region.\\
In their model they didn't include Gluonic interaction. Bicke11er, Birse and Wilets \citep{Bicke11erBirseandWilets, Bicke11erBirseandWilets1} have investigated a self-consistent one gluon exchange approximation with the consideration of Friedberg and Lee, but their work included an Abelian approximation. Also Haider and Liu \citep{HaiderandLiu, HaiderandLiu1} have also calculated hadron properties similarly, but using a step function for the dielectric function inside the bag. After it Dodd \citep{Dodd} proposed gluonic corrections to the models.\\
The model of Friedberg and Lee identifies the confining regions with a zero dielectric constant or infinite coupling constant. This model however also involves a fine tuning because the potential of the field $ \sigma $ has to have a minimum at the same place where the dielectric constant is zero.
More recently 't Hooft in his paper "Perturbative Confinement" \citep{tHooft2002} has put into question the idea that confinement must to be associated to a very big coupling constant . Several models has been given where the t' Hooft idea can be realized \citep{Kharzeev2008,Guendelman2003ib,Gaete2006xd}.\\
As we explained before in our case indeed the confinement will not be necessarily related to a very big coupling constant, in fact it will be related instead to the soliton behavior of the scalar field and to the space dependent of a dynamical coupling constant.\\
The article will be organized as follows. In section \ref{review} we review the work that was presented in \citep{GlobalSQED} .In section \ref{section abelian} we present another six approaches (all base on the same principles) in the Abelian case which give boundary conditions on a Dirac current or a scalar current and also consider a constraint on a magnetic moment current. The new approaches are related but different in details than the procedure in reference \citep{GlobalSQED}, and every one is interesting by its on right. Then in section \ref{section non abelian} we present four cases which gives boundary condition in non Abelian field in parallel to the cases of section \ref{section abelian} . In case one we generalize what we did at reference \citep{GlobalSQED} to non Abelian fields. In cases two three and four we generalize the example in section \ref{section abelian}. in section \ref{sec:The bubble scalar field} we will give an example  which can produce an additional realization of confinement.

\section{Review of Confining Boundary conditions in Abelian case }\label{review}
 In this chapter we review the work that we have done in ref \citep{GlobalSQED}. We will show that dynamical Coupling Constants can lead to confinement in Abelian field.
The dynamical Coupling Constants is dynamical mostly at the boundary of the confinement and outside the boundary.  The gauge coupling has a term of the form $g(\phi)j_\mu (A^{\mu} +\partial^{\mu}B)$ where $B$ is an auxiliary field and the current $j_\mu$ is the Dirac current.
Before studying the issue of dynamical gauging, we review how the $ B $ field can be used in a gauge theory playing the role of a scalar gauge field \citep{Scalargaugefield}\footnote{for previous, but less general treatments involving scalar gauge field see \citep{Guendelman2013,Stueckelberg,Guendelman1979}}. That can be used to define a new type of convariante derivative.
Starting with a complex scalar field we now gauge the phase symmetry of $\phi$ by introducing a real, scalar $B( x _\mu)$ and two types of covariant derivatives as
\begin{equation}
\label{cov-ab}
D ^A _\mu = \partial _\mu + i e A_\mu ~~~;~~~ D ^B _\mu = \partial _\mu + i e \partial _\mu B ~.  
\end{equation}  
The gauge transformation of the complex scalar, vector gauge field and scalar gauge field have the 
following gauge transformation
\begin{equation} \label{gauge-trans}
\phi \rightarrow e^{i e \Lambda} \phi ~~~;~~~ A_\mu \rightarrow A_\mu + \partial _\mu \Lambda ~~~;~~~
B \rightarrow B - \Lambda ~.
\end{equation} 
It is easy to see that terms like $D ^A _\mu \phi$ and $D ^B _\mu \phi$, will be
covariant under \ref{gauge-trans} that is they transform the same way as the scalar field $ \phi $ and their complex conjugates will transfor as $\phi^*$ does. Thus one can generate kinetic energy type terms like 
$(D ^A _\mu \phi) (D ^{A \mu} \phi)^*$, $(D ^B _\mu \phi) (D ^{B \mu} \phi)^*$, $(D ^A _\mu \phi) (D ^{B \mu} \phi)^*$,
and $(D ^B _\mu \phi) (D ^{A \mu} \phi)^*$. Unlike $A_\mu$ where one can add a gauge invariant 
kinetic term involving only $A_\mu$ (i.e. $F_{\mu \nu} F^{\mu \nu}$) this is apparently not possible to do for the
scalar gauge field $B$. However note that the term $A_\mu + \partial _\mu B$ is invariant under the
gauge field transformation alone (i.e. $A_\mu \rightarrow A_\mu + \partial _\mu \Lambda$ and 
$B \rightarrow B - \Lambda$). Thus one can add a term like $(A_\mu + \partial _\mu B)(A^\mu + \partial ^\mu B)$
to the Lagrangian which is invariant with respect to the gauge field part only of the gauge transformation
in \ref{gauge-trans}. This gauge invariant term will lead to both mass-like terms for the vector gauge
field and kinetic energy-like terms for the scalar gauge field. In total a general Lagrangian which respects 
the new gauge transformation and is a generalization of the usual gauge Lagrangian, which has 
the form
\begin{eqnarray} \label{u2}
& {\cal L} = c_1 D^A _\mu \phi (D ^{A \mu} \phi ) ^* + c_2 D^B _\mu \phi (D ^{B \mu} \phi )^* 
\nn & + c_3 D^A _\mu \phi (D ^{B \mu} \phi )^* 
 + c_4 D^B _\mu \phi (D ^{A \mu} \phi )^* - V(\phi) \nonumber \\ &
- \frac{1}{4} F_{\mu \nu} F^{\mu \nu} + c_5 (A_\mu + \partial _\mu B)(A^\mu + \partial ^\mu B)
\end{eqnarray}
where $c_i$'s are constants that should be fixed to get a physically acceptable Lagrangian where $ c_{3}=c^{*}_{4} $ and $ c_{1}\, , c_{2}\, , c_{5} $ are real.\\
At first glance one might conclude that $B(x)$ is not a physical field, it appears that one could "gauge" it away by taking $\Lambda = B(x)$ in \ref{gauge-trans}. However in the case of symmetry breaking when one introduces a complex charged scalar field that get expectation value which is not zero, one must be careful since this would imply that the gauge transformation of the field $\phi$ would be of the form $\phi \rightarrow e^{i e B} \phi$ i.e. the phase factor would be fixed by the gauge transformation of $B(x)$. In this situation one would no longer to able to use the usual unitary gauge transformation to eliminate the Goldstone boson in the case when one has spontaneous symmetry breaking. \\ Indeed in the case when there is spontaneous symmetry breaking, the physical gauge (the  generalization of the unitary gauge) is not  the gauge $B=0$, as discussed in \citep{Scalargaugefield}, it is a gauge where the scalar gauge field $B$ has to be taken proportional to the phase of the scalar field $ \theta $, with a proportionality constant that depends on the expectation value of the Higgs field acording to 
\begin{equation}
\label{physical gauge}
\theta = \frac{c_5 - a e^2 \rho_0 ^2}{e \rho_0 ^2(c_1 + a)} B ~. 
\end{equation}

Also, in general there are the three degrees of freedom of a massive vector field and the Higgs field, and therefore all together five degrees of freedom.

If there is no spontaneous symmetry breaking, fixing the gauge $B=0$  does not coincide with the gauge that allows us to display that the photon has two polarizations, this gauge being Coulomb gauge. This is true even if we do not add a gauge invariant mass term (possible given the existence of the $B$ field).
By fixing the Coulomb gauge, which will make the the photon manifestly having only two polarizations, we will have already exhausted the gauge freedom and cannot in general in addition require the gauge $B=0$. So, in Coulomb gauge where the photon will have two polarizations, the 
$B$ field and in addition the two other scalars, the real and imaginary part of $\phi$ all represent true degrees of freedom, so altogether we have five degrees of freedom, the same as the case displaying spontaneous symmetry breaking. If we add a gauge invariant mass term, even when there is no spontaneous symmetry breaking (the $c_5$ term), in the gauge $B=0$ we have three polarizations of the massive vector field and still the real and imaginary parts of the complex scalar field $\phi$, still five degrees of freedom altogether.

\subsection{Confining Boundary conditions from dynamical Coupling Constants in Abelian case}
We begin with Dirac field $ \psi $ and a real scalar field $ \phi $, with the action:
\begin{eqnarray}\label{Dirac:boundary}
& S=\int{\bar{\psi}(i\gamma^{\mu}\partial_{\mu}-m+e\gamma^{\mu}A_{\mu})\psi \,d^{4}x}
\nn & - \frac{1}{4} \int{F^{\mu\nu}F_{\mu\nu}\,d^{4}x} \nn & +\int d^{4}x [ g(\phi)\bar{\psi}\gamma^{\mu}\psi(A_{\mu}+\partial_{\mu}B)\nn & +\frac{1}{2}\partial_{\mu}\phi\partial^{\mu}\phi - V(\phi)]
\end{eqnarray}
The model is  invariant under local gauge transformations
\begin{equation} \label{GTGlobalQED2}
A^\mu \rightarrow A^\mu + \partial^\mu \Lambda \textrm{;   } \ \ \ 
B \rightarrow  B - \Lambda 
\end{equation}

\begin{equation}
\psi \rightarrow exp (ie\Lambda) \psi
\end{equation} 

The Noether current conservation law for global symmetry $ \psi \rightarrow e^{i\theta}\psi $, $\theta= constant$ is,

\begin{equation}
\partial_{\mu} j^{\mu}_{N}=(\partial_{\mu})(\frac{\partial \mathcal{L}}{\partial \psi_{,\mu}}\delta\psi) = \partial_{\mu}(\bar{\psi}\gamma^{\mu}\psi)=0
\end{equation}
The  gauge field equation, containing in the right hand side the current which  is the source of the gauge field is:
\begin{eqnarray}
\partial_{\mu}F^{\mu\nu}=(e+g(\phi))\bar{\psi}\gamma^{\nu}\psi = j^{\nu}_{Source}
\end{eqnarray}
By considering the divergence of the above equation, we obtain the additional conservation law:
\begin{eqnarray}\label{bag conservation law}
& \partial_{\mu}j^{\mu}_{Source}=\partial_{\mu}(g(\phi))\bar{\psi}\gamma^{\mu}\psi + g(\phi) \partial_{\mu}(\bar{\psi}\gamma^{\mu}\psi)\nn & = \partial_{\mu}(g(\phi))\bar{\psi}\gamma^{\mu}\psi = 0
\end{eqnarray}
If we have scalar potential $ V(\phi) $ with domain wall between two false vacuum state, than because of the transition of the scalar field on the domain wall $ \partial_{\mu}(g(\phi))=\frac{\partial g(\phi)}{\partial \phi}\partial_{\mu}\phi=\frac{\partial g(\phi)}{\partial \phi}n_{\mu}f \neq 0 $.
We must conclude that $ n_{\mu}(\bar{\psi}\gamma^{\mu}\psi)\mid_{x=domain\,wall}=0 $. This means that on the domain wall there is no communication between the two sector of the domain, which give a confinement.

\subsection{Confining Boundary conditions  holding at a specific region in a domain wall - Abelian case}\label{Confining Boundary conditions  holding at a specific region in a domain wall - Abelian case}
The constraint that we have gotten in the last subsection is too strong and non trivial and holds everywhere even if the gradients are small , but we wants to have constraint only in the region where the domain wall is located, so lets proceed with the same consideration as above but with coupling of the gauge field and the scalar field so the constraint will follow from an equation of motion and only at a specific location. We will see that if in the action we add an additional term (the $ 1/l_0 $ term):
\begin{eqnarray}\label{Dirac:boundary2}
& S=\int{\bar{\psi}(i\gamma^{\mu}\partial_{\mu}-m+e\gamma^{\mu}A_{\mu})\psi \,d^{4}x} 
\nn & - \frac{1}{4} \int{F^{\mu\nu}F_{\mu\nu}\,d^{4}x} \nn & +\int d^{4}x [ g(\phi)\bar{\psi}\gamma^{\mu}\psi(A_{\mu}+\partial_{\mu}B)+\frac{1}{2}\partial_{\mu}\phi\partial^{\mu}\phi - V(\phi) \nn & +\frac{1}{l_{0}}\partial_{\mu} \phi (A^{\mu} +\partial^{\mu}B)]
\end{eqnarray}
The model is  invariant under local gauge transformations as in equation \ref{GTGlobalQED2}.

The Noether current conservation law for global symmetry \\ $ \psi \rightarrow e^{i\theta}\psi $, $\theta= constant$ is,
\begin{equation}\label{Noether conservation law 2}
\partial_{\mu} j^{\mu}_{N}=(\partial_{\mu})(\frac{\partial \mathcal{L}}{\partial \psi_{,\mu}}\delta\psi) = \partial_{\mu}(\bar{\psi}\gamma^{\mu}\psi)=0
\end{equation}
The  gauge field equation, containing in the right hand side the current which  is the source of the gauge field is:
\begin{eqnarray}
\partial_{\mu}F^{\mu\nu}=(e+g(\phi))\bar{\psi}\gamma^{\nu}\psi +\frac{1}{l_{0}}\partial^{\nu}\phi = j^{\nu}_{Source}
\end{eqnarray}
We can see that we have additional long range term to the constraint \ref{bag conservation law}.
By considering the divergence of the above equation, we obtain the additional conservation law:
\begin{eqnarray}\label{bag conservation law 2}
& \partial_{\mu}j^{\mu}_{Source}=\partial_{\mu}(g(\phi))\bar{\psi}\gamma^{\mu}\psi + g(\phi) \partial_{\mu}(\bar{\psi}\gamma^{\mu}\psi) 
\nn &+\frac{1}{l_{0}}\partial^{\mu}\partial_{\mu}\phi 
 = \partial_{\mu}(g(\phi))\bar{\psi}\gamma^{\mu}\psi +\frac{1}{l_{0}}\partial^{\mu}\partial_{\mu}\phi = 0
\end{eqnarray}

The variation on the action by $ \phi $ gives:
\begin{eqnarray}\label{bag EOM phi}
& \partial^{\mu}\partial_{\mu}\phi + \frac{\partial V}{\partial \phi} - \frac{1}{l_{0}} \partial_{\mu}(A^{\mu}+\partial^{\mu}B) \nn & + \frac{\partial g(\phi)}{\partial \phi}\bar{\psi}\gamma^{\mu}\psi(A_{\mu}+\partial_{\mu}B)=0
\end{eqnarray}

Lets consider a scalar potential $ V(\phi) $ with domain wall between two false vacuum state $ V(\nu_{1}) $ and $ V(\nu_{2}) $, and statically solution. Than for finite energy solution we need to demand that $ \partial_{i}\phi(\pm \infty)=0 $ and $ \phi(\infty)=\nu_{1} $ and $ \phi(-\infty)=\nu_{2} $.

From Rolle's mathematical theorem, we must conclude that at some point of the transition of the scalar field on the domain wall we have that $ \partial_{i}\partial_{i}\phi=0 $ , so equation \ref{bag conservation law 2} on some point on the transition reads:
\begin{eqnarray}
\partial_{\mu}(g(\phi))\bar{\psi}\gamma^{\mu}\psi=0
\end{eqnarray}
 because on the point of the transition where $ \partial_{i}\partial_{i}\phi=0 $, $ \partial_{i}\phi \neq 0 $ than $ \partial_{\mu}(g(\phi))=\frac{\partial g(\phi)}{\partial \phi}\partial_{\mu}\phi=\frac{\partial g(\phi)}{\partial \phi}n_{\mu}f \neq 0 $.
\\So we must conclude that $ n_{\mu}(\bar{\psi}\gamma^{\mu}\psi)\mid_{x=domain\,wall}=0 $. This means that on the domain wall there is no communication between the two sector of the domain, which give a confinement. Also we can see that the coupling constant far from the domain wall is constant.

\section{Confining Boundary conditions in Abelian case - another approach.}\label{section abelian}
In this section we present new models which gives a parallel results as in ref \citep{GlobalSQED}. Every approach have different interpretation.
\subsection{Introduction of "dielectic potentials".}
We begin with the action
\begin{eqnarray}\label{Dirac:boundary02}
& S=\int{\lbrace\bar{\psi}(i\gamma^{\mu}\partial_{\mu} - m - \gamma^{\mu}A_{\mu})\psi + \mathcal{L}_{\phi}\rbrace \,d^{4}x} \nn & - \frac{1}{4}\int{D^{\mu\nu}D_{\mu\nu}\,d^{4}x}
\end{eqnarray}
where we define the dielectric field strength:
\begin{equation}\label{eq:dielectric field strength}
D_{\mu\nu}= \partial_{\mu}C_{\nu} - \partial_{\nu}C_{\mu}
\end{equation}
where $ C_{\mu} $ the dielectric potential is defined as:
\begin{equation}
C_{\mu} = (1+g(\phi))A_{\mu} + g(\phi)\partial_{\mu}B
\end{equation}
The Noether current conservation law for global symmetry $ \psi \rightarrow e^{i\theta}\psi $, $\theta= constant$ is,
\begin{equation}
\partial_{\mu} j^{\mu}_{N}=(\partial_{\mu})(\frac{\partial \mathcal{L}}{\partial \psi_{,\mu}}\delta\psi) = \partial_{\mu}(\bar{\psi}\gamma^{\mu}\psi)=0
\end{equation}

The model is  invariant under local gauge transformations
\begin{equation} \label{GTGlobalQED3}
A^\mu \rightarrow A^\mu + \partial^\mu \Lambda \textrm{;   } \ \ \ 
B \rightarrow  B - \Lambda 
\end{equation}

\begin{equation}
\psi \rightarrow exp (-i\Lambda) \psi
\end{equation} 

As we can check, under the transformations \ref{GTGlobalQED3}:
\begin{eqnarray}
C^\mu \rightarrow C^\mu + \partial^\mu \Lambda
\end{eqnarray}
and therefor $ D_{\mu\nu} $ is gauge invariant.
With this definition:
\begin{eqnarray}
& D_{\mu\nu} = (1+g(\phi))[\partial_{\mu}A_{\nu} - \partial_{\nu}A_{\mu}] \nn & + (A_{\nu}+\partial_{\nu}B)\partial_{\mu}g(\phi) - (A_{\mu}+\partial_{\mu}B)\partial_{\nu}g(\phi)
\end{eqnarray}
Variation of the action by $ A_{\mu} $ gives 
\begin{eqnarray}\label{abelian A eqaution}
\partial_{\mu}D^{\mu\nu} =\frac{\bar{\psi}\gamma^{\nu}\psi }{(1+g(\phi))} 
\end{eqnarray}
variation on the action \ref{Dirac:boundary02} by $ B $ gives:
\begin{eqnarray}\label{Boundery Abelian}
(\partial_{\nu}g(\phi))\partial_{\mu}D^{\mu\nu} = (\partial_{\nu}g(\phi)) \frac{\bar{\psi}\gamma^{\nu}\psi }{(1+g(\phi))} = 0 
\end{eqnarray}

where we have recognized that $ \partial_{\mu}(\bar{\psi}\gamma^{\mu}\psi) =0 $ and $ \partial_{\mu}D^{\mu\nu} =\frac{\bar{\psi}\gamma^{\nu}\psi }{(1+g(\phi))}  $.
Equation \ref{Boundery Abelian} tells us that, when the scalar field is changing then the Dirac field current will be zero in the direction of the change of the scalar field.
\\We can get the same result by derivative equation \ref{abelian A eqaution} so
\begin{eqnarray}\label{constraint: Dialectric potetial}
&\partial_{\nu}\partial_{\mu}D^{\mu\nu} =(\partial_{\nu}g(\phi))\frac{\bar{\psi}\gamma^{\nu}\psi }{(1+g(\phi))^{2}}  =0
\end{eqnarray}
Again the constraint that we have gotten is too strong and non trivial and holds everywhere even if the gradients are small , but we wants to have constraint only in the region where the domain wall is located, some can proceed with the same consideration as above \ref{Confining Boundary conditions  holding at a specific region in a domain wall - Abelian case} with adding in the action an additional term so the constraint will follow from an equation of motion and only at a specific location.

\subsection{Constraint from 4 point interaction of scalar field current and Dirac current.}

We begin with the action that have 4 point scalar and Dirac current interaction :
\begin{eqnarray}\label{Action:general phase control on Dirac2}
& S=\int \lbrace\bar{\psi}(i\gamma^{\mu}\partial_{\mu} - m - \gamma^{\mu}A_{\mu})\psi -\frac{1}{4} F^{\mu\nu}F_{\mu\nu}
\nn & + (\partial_{\mu}\phi^{*} + iA_{\mu}\phi^{*})(\partial^{\mu}\phi  - iA^{\mu}\phi ) - V(\phi^{*}\phi) \rbrace \,d^{4}x 
\nn & + \int{i\bar{\psi}\gamma^{\mu}\psi \,(\phi^{*} \partial_{\mu}\phi - \partial_{\mu}\phi^{*}\,\phi - 2i \phi^{*}\phi \, \partial_{\mu}B )\,d^{4}x}
\end{eqnarray}
Where 
\begin{eqnarray}
& F_{\mu\nu}=\partial_{\mu}A_{\nu}- \partial_{\nu}A_{\mu}
\end{eqnarray}
This expression is local gauge invariant anther:
\begin{eqnarray}\label{Gauge transformation partialB}
& \psi \rightarrow e^{-i\Lambda}\psi
\nn & \phi \rightarrow e^{i\Lambda}\phi
\nn & A_{\mu} \rightarrow A_{\mu} + \partial_{\mu}\Lambda
\nn & B \rightarrow B + \Lambda
\end{eqnarray}
The Noether current conservation law for global symmetry $ \phi \rightarrow e^{i\theta}\phi $, $\theta= constant$ is,:
\begin{eqnarray}\label{J(N phi)2}
J_{(N\phi) \mu}= \phi^{*}\partial_{\mu}\phi - (\partial_{\mu}\phi^{*}) \, \phi - i( 2 A_{\mu} + \bar{\psi}\gamma_{\mu}\psi)\phi^{*}\phi
\end{eqnarray}
and the Noether current conservation law for global symmetry $ \psi \rightarrow e^{i\theta}\psi $, $\theta= constant$ is,:
\begin{eqnarray}\label{J(N psi)2}
J^{\mu}_{(N \psi)}= i\bar{\psi}\gamma^{\mu}\psi
\end{eqnarray}
Variation by $ A_{\mu} $ gives the current:
\begin{eqnarray}\label{variation by A2}
\partial_{\nu}F^{\nu\mu} = \bar{\psi}\gamma^{\mu}\psi - i(\phi^{*}\partial_{\mu}\phi - (\partial_{\mu}\phi^{*}) \, \phi) - 2A_{\mu}\phi^{*}\phi
\end{eqnarray}
Variation by $ B $ and using equation \ref{J(N psi)2} gives again the constraint:
\begin{eqnarray}\label{constraint:4 point interaction2}
\bar{\psi}\gamma^{\mu}\psi \, \partial_{\mu}(\phi^{*}\phi) = 0
\end{eqnarray}
We can achieve the same constraint with different theory, where $\partial_{\mu} B $ is replaced by the vector field $ B_{\mu} $. 
Now we considered the action:
\begin{eqnarray}\label{Action:general phase control on Dirac}
& S=\int \lbrace\bar{\psi}(i\gamma^{\mu}\partial_{\mu} - m - \gamma^{\mu}A_{\mu})\psi -\frac{1}{4} F^{\mu\nu}F_{\mu\nu}
\nn & + (\partial_{\mu}\phi^{*} + iA_{\mu}\phi^{*})(\partial^{\mu}\phi  - iA^{\mu}\phi ) - V(\phi^{*}\phi) \rbrace \,d^{4}x 
\nn & + \int{i\bar{\psi}\gamma^{\mu}\psi \,(\phi^{*} \partial_{\mu}\phi - \partial_{\mu}\phi^{*}\,\phi - 2i \phi^{*}\phi B_{\mu} )\,d^{4}x}
\nn & -\frac{1}{4}\,\int{B_{\mu\nu}B^{\mu\nu}}\,d^{4}x -\frac{\lambda}{2}\int {F_{\mu\nu}B^{\mu\nu}}\,d^{4}x
\end{eqnarray}
Where 
\begin{eqnarray}
& F_{\mu\nu}=\partial_{\mu}A_{\nu}- \partial_{\nu}A_{\mu}
\nn & B_{\mu\nu}=\partial_{\mu}B_{\nu}- \partial_{\nu}B_{\mu}
\end{eqnarray}
This expression is local gauge invariant anther:
\begin{eqnarray}\label{Gauge transformation Bmu}
& \psi \rightarrow e^{-i\Lambda}\psi
\nn & \phi \rightarrow e^{i\Lambda}\phi
\nn & A_{\mu} \rightarrow A_{\mu} + \partial_{\mu}\Lambda
\nn & B_{\mu} \rightarrow B_{\mu} + \partial_{\mu}\Lambda
\end{eqnarray}
The Noether current conservation law for global symmetry $ \phi \rightarrow e^{i\theta}\phi $, $\theta= constant$ is,
\begin{eqnarray}\label{J(N phi)}
J_{(N\phi) \mu}= \phi^{*}\partial_{\mu}\phi - (\partial_{\mu}\phi^{*}) \, \phi - i( 2 A_{\mu} + \bar{\psi}\gamma_{\mu}\psi)\phi^{*}\phi
\end{eqnarray}
and the Noether current conservation law for global symmetry $ \psi \rightarrow e^{i\theta}\psi $, $\theta= constant$ is,
\begin{eqnarray}\label{J(N psi)}
J_{(N \psi)\mu}= i\bar{\psi}\gamma^{\mu}\psi
\end{eqnarray}
Variation by $ A_{\mu} $ gives the current:
\begin{eqnarray}\label{variation by A}
&\partial_{\nu}F^{\nu\mu} + \lambda \partial_{\nu}B^{\nu\mu} = \bar{\psi}\gamma^{\mu}\psi - i(\phi^{*}\partial_{\mu}\phi - (\partial_{\mu}\phi^{*}) \, \phi)
\nn & - 2A_{\mu}\phi^{*}\phi
\end{eqnarray}
Variation by $ B_{\mu} $ gives:
\begin{eqnarray}\label{variation by B}
\partial_{\nu}B^{\nu\mu} + \lambda \partial_{\nu}F^{\nu\mu} = - 2 \bar{\psi}\gamma^{\mu}\psi \phi^{*}\phi
\end{eqnarray}
From equation \ref{J(N psi)} and \ref{variation by B} we have again the constraint:
\begin{eqnarray}\label{constraint:4 point interaction}
\bar{\psi}\gamma^{\mu}\psi \, \partial_{\mu}(\phi^{*}\phi) = 0
\end{eqnarray}
\subsection{Derivation of the M.I.T confining boundary conditions for gauge fields}
We begin with an action that has dynamical electric coupling constant and auxiliary field $ B $:
\begin{eqnarray}\label{action: electric and magnetic field}
& S= \int \bar{\psi}(i\gamma^{\mu}\partial_{\mu} - m - g(\phi)\gamma^{\mu}A_{\mu}
\nn & + (\partial_{\mu}g(\phi))\gamma^{\mu}B)\psi \, d^{4}x  - \frac{1}{4} \int g(\phi)F^{\mu\nu}F_{\mu\nu} + \mathcal{L}_{\phi}
\end{eqnarray}
where:
\begin{eqnarray}
F_{\mu\nu} = \partial_{\mu}A_{\nu} - \partial_{\nu}A_{\mu}
\end{eqnarray}
and the action is local gauge invariant anther:
\begin{eqnarray}
& \psi \rightarrow e^{-ig(\phi)\Lambda}\psi
\nn & A_{\mu} \rightarrow A_{\mu} + \partial_{\mu}\Lambda
\nn & B \rightarrow B - \Lambda
\end{eqnarray}
The Noether current conservation law for global symmetry $ \psi \rightarrow e^{i\theta}\psi $, $\theta= constant$ is:
\begin{eqnarray}
J_{N} = i\bar{\psi}\gamma^{\mu}\psi
\end{eqnarray}
variation by $ A_{\mu} $ gives:
\begin{eqnarray}\label{F constariant}
(\partial_{\mu}g(\phi))F^{\mu\nu} + g(\phi)\partial_{\mu}F^{\mu\nu} = g(\phi)\bar{\psi }\gamma^{\mu}\psi
\end{eqnarray}
We can see that if $ g(\phi)=0 $ and also $ \partial_{\mu}g(\phi)\neq 0 $ then we can have the constraint:
\begin{eqnarray}\label{constraint: electric field 1}
(\partial_{\mu}g(\phi))F^{\mu\nu} = 0
\end{eqnarray}
This corresponds to the M.I.T boundary conditions for confinement of gauge fields.
Furthermore, by taking divergence of equation \ref{F constariant} or alternatively by variation of the action with respect to the field $ B $ we get:
\begin{eqnarray}
(\partial_{\mu}g(\phi))\bar{\psi}\gamma^{\mu}\psi = 0
\end{eqnarray} 
So in this case we can have two constraint, one on the current and the second on the electric and magnetic field.
\subsection{Model that gives M.I.T confinement condition for gauge fields and scalar current constraint}
With the same procedure like the previous subsection we can make also constraint on the scalar current, so we begin with the action:
\begin{eqnarray}\label{action: electric and magnetic field 2}
& S= \int \lbrace (\partial_{\mu}\phi^{*} + ig(\sigma)A_{\mu}\phi^{*} - i(\partial_{\mu}g(\sigma))B\phi^{*})
\nn &(\partial^{\mu}\phi  - ig(\sigma)A^{\mu}\phi + i(\partial^{\mu}g(\sigma))B\phi )
\nn & - V(\phi^{*}\phi) - \frac{g(\sigma)}{4}F^{\mu\nu}F_{\mu\nu} +\mathcal{L}_{\sigma} \rbrace \,d^{4}x 
\end{eqnarray}
where:
\begin{eqnarray}
F_{\mu\nu} = \partial_{\mu}A_{\nu} - \partial_{\nu}A_{\mu}
\end{eqnarray}
and the action is local gauge invariant under:
\begin{eqnarray}
& \phi \rightarrow e^{-ig(\sigma)\Lambda}\phi
\nn & A_{\mu} \rightarrow A_{\mu} - \partial_{\mu}\Lambda
\nn & B \rightarrow B + \Lambda
\end{eqnarray}
The Noether current conservation law for global symmetry $ \phi \rightarrow e^{i\theta}\phi $, $\theta= constant$ is:
\begin{eqnarray}
& J^{\mu}_{N} = \phi^{*} \partial^{\mu}\phi - \partial^{\mu}\phi^{*}\,\phi
\nn & - 2i A^{\mu} \phi^{*}\phi + 2i(\partial^{\mu}g(\sigma))B \phi^{*}\phi
\end{eqnarray}
Variation by $ A_{\mu} $ gives:
\begin{eqnarray}
(\partial_{\mu}g(\sigma))F^{\mu\nu} + g(\sigma)\partial_{\mu}F^{\mu\nu} = -ig(\sigma)J^{\nu}_{N}
\end{eqnarray}
We can see that if $ g(\phi)=0 $ and also $ \partial_{\mu}g(\phi)\neq 0 $ then we can have the constraint:
\begin{eqnarray}\label{constraint: electric field 2}
(\partial_{\mu}g(\phi))F^{\mu\nu} = 0
\end{eqnarray}
Variation by $ B $ gives the second constraint on the scalar field current:
\begin{eqnarray}\label{constraint scalar current}
& i(\partial_{\mu}g(\sigma))[\phi^{*} \partial_{\mu}\phi - \partial_{\mu}\phi^{*}\,\phi 
\nn & - 2i A_{\mu} \phi^{*}\phi + 2i(\partial_{\mu}g(\sigma))B \phi^{*}\phi]=
\nn & i(\partial_{\mu}g(\sigma))J^{\mu}_{N} = 0
\end{eqnarray}
\subsection{Coupling to magnetic moment type interaction:}
We begin with action that have spinor interaction, that interaction have variant coupling constant.
Mostly spin interaction is in the form $ \bar{\psi}\sigma^{\mu\nu}\psi F_{\mu\nu} $ but by integration by part it can deform to be $ \partial_{\mu}(\bar{\psi}\sigma^{\mu\nu}\psi) A_{\mu} $. We will show that if we have variant coupling constant to this interaction and exalary field, we can get constraint on the spinor.
\begin{eqnarray}\label{Action: spinor1 }
& S=\int{\bar{\psi}(i\gamma^{\mu}\partial_{\mu}-m +  e\bar{\psi}\gamma^{\mu}A_{\mu}\psi)\psi \,d^{4}x}  \nn & - \frac{1}{4} \int{F^{\mu\nu}F_{\mu\nu}\,d^{4}x}
+\int d^{4}x [ \mu(\phi)\partial_{\nu}(\bar{\psi}\sigma^{\mu \nu}\psi)(A_{\mu}+\partial_{\mu}B)
\nn & +\frac{1}{2}\partial_{\mu}\phi\partial^{\mu}\phi - V(\phi)]
\end{eqnarray} 
where $ \sigma^{\mu\nu}=\frac{i}{4}[\gamma^{\mu} , \gamma^{\nu}] $.
The model is  invariant under local gauge transformations
\begin{equation} \label{gauge transformations:spinor 1}
A^\mu \rightarrow A^\mu + \partial^\mu \Lambda \textrm{;   } \ \ \ 
B \rightarrow  B - \Lambda 
\end{equation}

\begin{equation}
\psi \rightarrow exp (ie\Lambda) \psi
\end{equation}
The Nuether current is:
\begin{eqnarray}\label{nuther current: spinor 1}
J_{N}^{\mu}= i\bar{\psi}\gamma^{\mu}\psi
\end{eqnarray}
Variation by $ A_{\mu} $ gives:
\begin{eqnarray}\label{Variation A:spinor 1}
\partial_{\mu}F^{\mu\nu} = -e\bar{\psi}\gamma^{\nu}\psi - \mu(\phi)\partial_{\mu}(\bar{\psi}\sigma^{\mu \nu}\psi)=j^{\nu}_{A}
\end{eqnarray}
from equation \ref{nuther current: spinor 1} and \ref{Variation A:spinor 1} or parallel by variation by $ B $ on the action we get the constraint on the spinor:
\begin{eqnarray}\label{constraint:spinor}
(\partial_{\mu} \mu(\phi))\partial_{\nu}(\bar{\psi}\sigma^{\mu\nu}\psi) = 0
\end{eqnarray}

\subsection{Dirac spinor and scalar current interaction.}
We going to show that the same constraint can be follow as in the previews subsection with current spin interaction. We begin with the action:
\begin{eqnarray}\label{Action:Dirac spinor 2 }
& S=\int \lbrace\bar{\psi}(i\gamma^{\mu}\partial_{\mu} - m - \gamma^{\mu}A_{\mu})\psi -\frac{1}{4} F^{\mu\nu}F_{\mu\nu}
\nn & + (\partial_{\mu}\phi^{*} + iA_{\mu}\phi^{*})(\partial^{\mu}\phi  - iA^{\mu}\phi ) - V(\phi^{*}\phi) \rbrace \,d^{4}x 
\nn & + \int{i\partial_{\nu}(\bar{\psi}\sigma^{\mu\nu}\psi) \,(\phi^{*} \partial_{\mu}\phi - \partial_{\mu}\phi^{*}\,\phi - 2i \phi^{*}\phi B_{\mu} )\,d^{4}x}
\nn & -\frac{1}{4}\,\int{B_{\mu\nu}B^{\mu\nu}}\,d^{4}x - -\frac{\lambda}{2}\,\int{B_{\mu\nu}F^{\mu\nu}}\,d^{4}x 
\end{eqnarray}
Where 
\begin{eqnarray}
& F_{\mu\nu}=\partial_{\mu}A_{\nu}- \partial_{\nu}A_{\mu}
\nn & B_{\mu\nu}=\partial_{\mu}B_{\nu}- \partial_{\nu}B_{\mu}
\end{eqnarray}
This expression is gauge invariant anther:
\begin{eqnarray}\label{Gauge transformation Bmu spinor}
& \psi \rightarrow e^{-i\Lambda}\psi
\nn & \phi \rightarrow e^{i\Lambda}\phi
\nn & A_{\mu} \rightarrow A_{\mu} + \partial_{\mu}\Lambda
\nn & B_{\mu} \rightarrow B_{\mu} + \partial_{\mu}\Lambda
\end{eqnarray}
The Nuether current is:
\begin{eqnarray}\label{J(N phi) spinor}
& J_{(N\phi) \mu}= \phi^{*}\partial_{\mu}\phi - (\partial_{\mu}\phi^{*}) \, \phi
\nn & - i( 2 A_{\mu} + \partial_{\nu}(\bar{\psi}\sigma^{\mu\nu}\psi))\phi^{*}\phi
\end{eqnarray}
\begin{eqnarray}\label{J(N psi)spinor}
J_{(N \psi)\mu}= i\bar{\psi}\gamma^{\mu}\psi
\end{eqnarray}
Variation by $ A_{\mu} $ gives the current:
\begin{eqnarray}\label{variation by A spinor}
& \partial_{\nu}F^{\nu\mu} + \lambda \partial_{\nu}B^{\nu\mu} =
\nn & \bar{\psi}\gamma^{\mu}\psi - i(\phi^{*}\partial_{\mu}\phi - (\partial_{\mu}\phi^{*}) \, \phi) - 2A_{\mu}\phi^{*}\phi
\end{eqnarray}
Variation by $ B_{\mu} $ gives:
\begin{eqnarray}\label{variation by B spinor}
\partial_{\nu}B^{\nu\mu} + \lambda \partial_{\nu}F^{\nu\mu} = - 2 \partial_{\nu}(\bar{\psi}\sigma^{\mu\nu}\psi) \phi^{*}\phi
\end{eqnarray}
From equation \ref{variation by B spinor} we have again the constraint:
\begin{eqnarray}\label{constraint:spinor 2}
\partial_{\mu}\partial_{\nu}B^{\nu\mu} + \lambda \partial_{\mu}\partial_{\nu}F^{\nu\mu} = - 2 \partial_{\nu}(\bar{\psi}\sigma^{\mu\nu}\psi)\partial_{\mu}( \phi^{*}\phi) = 0 
\end{eqnarray}
where we used the fact that $\partial_{\mu} \partial_{\nu}(\bar{\psi}\sigma^{\mu\nu}\psi)=0 $
\subsection{Consistent models with time dependent fine structure constant }
The formalism developed here provides the possibility of formulating a consistent formalism where the effective electric charge can change with space and time such possibility have been considered in cosmological contexts.
Many papers have been published on the subject of the variation of the fine structure constant.
There are some clues that show that the structure constant has been slightly variable, although this is not generally agreed.
Bekenstein \citep{Bekenstein} has shown a different approach to formulate consistently a theory with a variable coupling constant. The Oklo natural geological fission reactor has lead to a measurement that some claim it implies the structure constant has changed by a small amount of the order of $ \frac{\dot{\alpha}}{\alpha} \approx 1\times 10^{-7} $ \citep{Uzan}.
\section{Confining boundary conditions in $ SU(N) $ non Abelian case}\label{section non abelian}
In the case of non Abelian gauge fields, we can also obtain the constraint boundary conditions in a similar way.
First by adding an interaction between current and the gauge field of the form (introduction the dynamical coupling constant $ g(\phi) $ ). $  g(\phi)\bar{\psi}\gamma^{\mu}( T^{a}A^{a}_{\mu} + \frac{i}{\zeta} U(\theta)\partial_{\mu}U^{-1}(\theta))\psi $ where the unitary matrix $ U $ corresponds to the non Abelian generalization of the $ B $ field of Abelian case.
The second possibility is to introduce $ g(\phi) $ and $ U $ in a dielectric field strengths $ D_{\mu\nu} $. However, as opposed to the Abelian case the constraint takes place naturally at a certain value of $ \phi $, not every where, so the problem of the " too strong constraint" does not a pear in the non Abelian case. Furthermore we can see that also the 4 point interaction between scalar current and Dirac current in non Abelian field produce constraint on the current. Also like in the Abelian case the Spinor constraint can be produce.
Now we going to cover the possible non Abelian case for each of the above example.
\subsection{Case 1: Constraint from Dynamical Couplings in the Non Abelian Theory}
Let us proceed with the same consideration but with Dirac field $ \psi $ and real scalar field $ \phi $, with the action that uses the gauge covariant combination\\ 
$ T^{a}A^{a}_{\mu} + \frac{i}{\zeta} U(\theta)\partial_{\mu}U^{-1}(\theta)$ which was used by Cornwall to construct gauge invariant mass terms \citep{Cornwall} but which we use here to couple to a current and to produce a dynamical coupling between the gauge fields and the fermions,
\begin{eqnarray}\label{Dirac:boundary20}
& S=\int{\bar{\psi}(i\gamma^{\mu}\partial_{\mu}-m- \zeta \gamma^{\mu}T^{a}A^{a}_{\mu})\psi \,d^{4}x}
\nn & - \frac{1}{4} \int{F^{a\mu\nu}F^{a}_{\mu\nu}\,d^{4}x} 
\nn & - \int{d^{4}x[ g(\phi)\bar{\psi}\gamma^{\mu}( T^{a}A^{a}_{\mu} + \frac{i}{\zeta} U(\theta)\partial_{\mu}U^{-1}(\theta))\psi]} 
\nn & + S_{\phi}
\end{eqnarray}
where $ T^{a} $ are the generators of $ SU(N) $, where $ T^{a} $ are normilazed as $ Tr(T^{a}T^{b}) = N\delta^{a b} $.
This action is invariant under gauge transformation that also involve the transformation of $U$ 
\begin{equation}\label{Utransf-Cornwalleq 1}
U \rightarrow V U
\end{equation}
so:
\begin{eqnarray}
& \frac{i}{\zeta} U(\theta)\partial_{\mu}U^{-1}(\theta) \rightarrow V(\frac{i}{\zeta} U(\theta)\partial_{\mu}U^{-1}(\theta)) V^{-1} 
\nn & + \frac{i}{\zeta} V \partial_{\mu}V^{-1}
\end{eqnarray}
while $T^{a}A^{a}_{\mu}$ transforms as
\begin{equation}\label{Atransf-Cornwalleq 1}
 T^{a}A^{a}_{\mu} \rightarrow V  T^{a}A^{a}_{\mu}V^{-1} - \frac{i}{\zeta} V \partial_{\mu}V^{-1} 
\end{equation}
The $ F^{a}_{\mu\nu} $ is defined:
\begin{equation}
F^{a}_{\mu\nu}=\partial_{\mu}A^{a}_{\nu} - \partial_{\nu}A^{a}_{\mu} + \zeta f^{a b c}A^{b}_{\mu} A^{c}_{\nu}
\end{equation}
Variation on the action by $ A^{a}_{\mu} $ gives:
\begin{equation}\label{EOM current variation A case 1}
D_{\mu}F^{a}_{\mu\nu}=(\zeta + g(\phi))\bar{\psi}T^{a}\gamma^{\nu}\psi
\end{equation}
Variation by $ U $ gives (where we used the fact\\ $\frac{i}{\zeta} U\partial_{\mu}U^{-1} = \frac{i}{N\zeta} T^{a} Tr(T^{a}U\partial_{\mu}U^{-1}) $:
\begin{eqnarray}
&\int[ -  \frac{i}{\zeta} g(\phi)\bar{\psi}T^{a}\gamma^{\mu}\delta (Tr(T^{a}U\partial_{\mu}U^{-1}))\psi] d^{4}x
\nn & = \int [ -  \frac{i}{\zeta} g(\phi)\bar{\psi}T^{a}\gamma^{\mu} Tr(T^{a}\delta(U\partial_{\mu}U^{-1}))\psi ]d^{4}x
\nn & =\int \lbrace -  \frac{i}{\zeta} g(\phi)\bar{\psi}T^{a}\gamma^{\mu}[Tr(T^{a} \delta(U)\partial_{\mu}U^{-1})
\nn & + Tr(T^{a} U\delta(\partial_{\mu}U^{-1}))]\psi \rbrace d^{4}x 
\nn & =\int \lbrace -  \frac{i}{\zeta} g(\phi)\bar{\psi}T^{a}\gamma^{\mu}[Tr(T^{a} (\delta U)\partial_{\mu}U^{-1})
\nn & - Tr(T^{a} \partial_{\mu} U\delta(U^{-1}))]\psi 
\nn & +  \frac{i}{\zeta} (\partial_{\mu}g(\phi))\bar{\psi}T^{a}\gamma^{\mu}Tr(T^{a}U\delta U^{-1})\psi
\nn & +  \frac{i}{\zeta} g(\phi)\partial_{\mu}(\bar{\psi}T^{a}\gamma^{\mu}Tr(T^{a}U\delta U^{-1})\psi) \rbrace d^{4}x = 0
\end{eqnarray}
Using the fact that
\begin{eqnarray}
& Tr(T^{a} \delta(U)\partial_{\mu}U^{-1}) - Tr(T^{a} \partial_{\mu}U\delta(U^{-1})) 
\nn & = - Tr([T^{a}, \partial_{\mu} U ]\delta U^{-1}])
\end{eqnarray}
we obtain that the equation of motion is
\begin{eqnarray}\label{case 1 bound U}
& Tr(T^{a}U\delta U^{-1})(\partial_{\mu}g(\phi))\bar{\psi}T^{a}\gamma^{\mu}\psi 
\nn & + g(\phi)[Tr(T^{a}U\delta U^{-1})\partial_{\mu}(\bar{\psi}T^{a}\gamma^{\mu}\psi)
\nn & + \bar{\psi}T^{a} Tr([T^{a}, \partial_{\mu} U ]\delta U^{-1}])\gamma^{\mu}\psi ] = 0
\end{eqnarray}

We can see that if $ g(\phi)=0 $ and $ \partial_{\mu}g(\phi)\neq 0 $ then $ \partial_{\mu}g(\phi)\bar{\psi}T^{a}\gamma^{\mu}\psi = 0 $. 
\\Furthermore, if we perform a covariant divergence on equation \ref{EOM current variation A case 1} we get:
\begin{eqnarray}\label{case 1 bound A}
& (\partial_{\mu}g(\phi))\bar{\psi}T^{a}\gamma^{\mu}\psi + (\zeta +g(\phi))[\partial_{\mu}(\bar{\psi}T^{a}\gamma^{\mu}\psi)
\nn & +i \zeta f^{a b c}\bar{\psi} A^{b}_{\mu} T^{c} \gamma^{\mu}\psi ] = 0
\end{eqnarray}

We can see that if $ g(\phi)= - \zeta $ and $ \partial_{\mu}g(\phi)\neq 0 $ then $ \partial_{\mu}g(\phi)\bar{\psi}T^{a}\gamma^{\mu}\psi = 0 $.
Not like in the simplest case of the Abelian symmetry where we got a too strong constraint such that we had to add a term to the action so that the constraint will be just at specific location (see subsection \ref{Confining Boundary conditions  holding at a specific region in a domain wall - Abelian case}). In the non Abelian case we get that the constraint is in a very specific location automatically.
In section \ref{sec:The bubble scalar field} we will see that from those consideration we can obtain a bubble formation.

\subsection{Case 2: Constraint using dielectric field in the non Abelian theory}
We start with the action:
\begin{eqnarray}\label{Dirac:boundary non Abilean}
& S=\int{\lbrace\bar{\psi}(i\gamma^{\mu}\partial_{\mu}-m - \zeta \gamma^{\mu}T^{a}A^{a}_{\mu})\psi \rbrace \,d^{4}x}
\nn & - \frac{1}{4}\int{D^{a\mu\nu}D^{a}_{\mu\nu}\,d^{4}x} + S_{\phi}
\end{eqnarray}
where:
\begin{eqnarray}
D^{a}_{\mu\nu}=\partial_{\mu}C^{a}_{\nu} - \partial_{\nu}C^{a}_{\mu} + \zeta f^{a b c} C^{b}_{\mu}C^{c}_{\nu}
\end{eqnarray}
where:
\begin{eqnarray}
&C_{\mu}= (1+g(\phi))T^{a}A^{a}_{\mu} +  \frac{i}{\zeta} g(\phi) U\partial_{\mu}U^{-1} = \nn &
(1+g(\phi))T^{a}A^{a}_{\mu} +  \frac{i}{\zeta} g(\phi)T^{a}\Tr(T^{a}U\partial_{\mu}U^{-1})
\end{eqnarray}
Under gauge transformation $U$ transform as  
\begin{equation}
U \rightarrow V U
\end{equation}
while $T^{a}A^{a}_{\mu}$ transforms as
\begin{equation}\label{Atransf-Cornwalleq 2}
g T^{a}A^{a}_{\mu} \rightarrow V g T^{a}A^{a}_{\mu}V^{-1} -  \frac{i}{\zeta} V \partial_{\mu}V^{-1} 
\end{equation}
and
\begin{equation}
C_{\mu} \rightarrow V C_{\mu} V^{-1} - \frac{i}{\zeta} V \partial_{\mu}V^{-1}
\end{equation}
while $ D^{a}_{\mu\nu} $ transforms co-variantly  as 
\begin{equation}
T^{a} D^{a}_{\mu\nu} \rightarrow V T^{a}D^{a}_{\mu\nu} V^{-1}
\end{equation}
Now $ D^{a}_{\mu\nu} $ will be:
\begin{eqnarray}
& D^{a}_{\mu\nu}=(1+g(\phi))(\partial_{\mu}A^{a}_{\nu}-\partial_{\nu}A^{a}_{\mu})
\nn &+(A^{a}_{\nu} +  \frac{i}{\zeta} \tran)\partial_{\mu}g(\phi)
\nn & - (A^{a}_{\mu} +  \frac{i}{\zeta} \tram)\partial_{\nu}g(\phi)
\nn & +  \frac{i}{\zeta} g(\phi)[Tr(T^{a}\partial_{\mu}U\partial_{\nu}U^{-1}) - Tr(T^{a}\partial_{\nu}U\partial_{\mu}U^{-1}]
\nn & + \zeta f^{a b c}[(1+g(\phi))A^{b}_{\mu}+  \frac{i}{\zeta} g(\phi)\trbm]* 
\nn & [(1+g(\phi))A^{c}_{\nu} +  \frac{i}{\zeta} g(\phi)\trcn]
\end{eqnarray}

Variation of the action \ref{Dirac:boundary non Abilean} with respect to $ A^{a}_{\mu} $ gives:
\begin{eqnarray}\label{BagNA variation A 1}
& \partial_{\mu}D^{a \mu\nu}+\zeta f^{a b c} [(1 + g(\phi))A^{b}_{\mu} \nn & +  i g(\phi)\Tr(T^{b}U\partial_{\mu}U^{-1}) ]D^{c \mu\nu} =
\nn & D^{(\textbf{C})}_{\mu}D^{a\mu\nu} =\frac{\zeta \bar{\psi}T^{a}\gamma^{\nu}\psi}{1+g(\phi)}
\end{eqnarray}
Variation of the action \ref{Dirac:boundary non Abilean} with respect to $ U $ gives:
\begin{eqnarray}\label{BagNA Boundary}
& Tr(T^{a}U\delta U^{-1}) \partial_{\mu}g(\phi)D^{(\textbf{C})}_{\nu}D^{\mu\nu a}
\nn & + Tr(T^{a}U\delta U^{-1}) g(\phi)[ \partial_{\mu}(D^{(\textbf{C})}_{\nu}D^{\mu\nu a}) 
\nn & +  Tr([T^{a}, \partial_{\mu} U ]\delta U^{-1}])D^{(\textbf{C})}_{\nu}D^{\mu\nu a}] = 0
\end{eqnarray}
We can notice in equation \ref{BagNA Boundary} that, if $ \partial_{\mu}g(\phi)\neq 0 $ and $ g(\phi)=0 $ then
\begin{eqnarray}
(\partial_{\mu}g(\phi)) D^{(\textbf{C})}_{\mu}D^{a \mu\nu} = 0
\end{eqnarray}
and 
\begin{equation}
(\partial_{\mu}g(\phi))\bar{\psi}T^{a}\gamma^{\mu}\psi = 0
\end{equation}

\subsection{Case 3: Constraint from Dirac and scalar current interaction.}
We begin with the action:
\begin{eqnarray}\label{Action:general phase control on Dirac non abelian}
& S=\int \lbrace\bar{\psi}(i\gamma^{\mu}\partial_{\mu} - m -\zeta \gamma^{\mu}T^{a}A^{a}_{\mu})\psi -\frac{1}{4} F^{a\mu\nu}F^{a}_{\mu\nu}
\nn & + (\partial_{\mu}\phi^{+} + i\zeta \phi^{+} T^{a}A^{a}_{\mu})(\partial^{\mu}\phi  - i\zeta T^{a}A^{a\mu}\phi ) - V(\phi^{+}\phi)  
\nn & +  i\bar{\psi}T^{a}\gamma^{\mu}\psi \,(\phi^{+} T^{a}\partial_{\mu}\phi - \partial_{\mu}\phi^{+}\,T^{a}\phi 
\nn &- 2\zeta i\epsilon^{abc} \phi^{+} T^{b} B^{c}_{\mu} \phi ) -\frac{1}{4}\,B^{a}_{\mu\nu}B^{a\mu\nu}
\nn & - \frac{\lambda}{2}\,B^{a}_{\mu\nu}F^{a\mu\nu} \rbrace \,d^{4}x
\end{eqnarray}
where:
\begin{equation}
F^{a}_{\mu\nu}=\partial_{\mu}A^{a}_{\nu} - \partial_{\nu}A^{a}_{\mu} + \zeta f^{a b c}A^{b}_{\mu} A^{c}_{\nu}
\end{equation}
and
\begin{equation}
B^{a}_{\mu\nu}=\partial_{\mu}B^{a}_{\nu} - \partial_{\nu}B^{a}_{\mu} + \zeta f^{a b c}B^{b}_{\mu} B^{c}_{\nu}
\end{equation}
The action is gauge invariant under:
\begin{eqnarray}
&\psi \rightarrow V\psi 
\nn &  T^{a}A^{a}_{\mu} \rightarrow V  T^{a}A^{a}_{\mu}V^{-1} - \frac{i}{\zeta} V \partial_{\mu}V^{-1}
\nn &  T^{a}B^{a}_{\mu} \rightarrow V  T^{a}B^{a}_{\mu}V^{-1} - \frac{i}{\zeta} V \partial_{\mu}V^{-1}
\end{eqnarray}
Variation by $ A^{a}_{\mu} $ gives:
\begin{eqnarray}
& D^{(\textbf{A})}_{\mu}F^{a\mu\nu} + \lambda D^{(\textbf{A})}_{\mu}B^{a\mu\nu} =
\nn & \zeta\bar{\psi}T^{a}\gamma^{\nu}\psi - i\zeta(\phi^{+} T^{a}\partial^{\nu}\phi - \partial^{\nu}\phi^{+}\,T^{a}\phi 
\nn &- 2\zeta i\epsilon^{abc} \phi^{+} T^{b} A^{c\nu} \phi )
\end{eqnarray}
where:
\begin{eqnarray}
D^{(\textbf{A})}_{\mu} = \partial_{\mu} - i\zeta T^{a}A^{a}_{\mu}
\end{eqnarray}
Variation by $ B_{\mu} $ gives:
\begin{eqnarray}
D^{(\textbf{B})}_{\mu}B^{a\mu\nu} + \lambda D^{(\textbf{B})}_{\mu}F^{a\mu\nu} = 2 \zeta \epsilon^{abc} \phi^{+} T^{b} \bar{\psi}\gamma^{\mu}T^{c}\psi \phi
\end{eqnarray}
where:
\begin{eqnarray}
D^{(\textbf{B})}_{\mu} = \partial_{\mu} - i\zeta T^{a}B^{a}_{\mu}
\end{eqnarray}
second darivation on the last equation gives:
\begin{eqnarray}
& D^{(\textbf{B})}_{\nu}D^{(\textbf{B})}_{\mu}B^{a\mu\nu} + \lambda D^{(\textbf{B})}_{\nu}D^{(\textbf{B})}_{\mu}F^{a\mu\nu} =
\nn & 2 \zeta\partial_{\nu}(\phi^{+}\phi) \epsilon^{abc} T^{b} \bar{\psi}\gamma^{\mu}T^{c}\psi + 
\nn & 2\zeta \phi^{+}\phi \epsilon^{abc}T^{b} [\partial_{\nu}(\bar{\psi}T^{c}\psi) - i\zeta \epsilon ^{cdf}B^{d}_{\nu}\bar{\psi} T^{f}\gamma^{\nu}\psi] = 0
\end{eqnarray}
We can see that if $ \phi^{+}\phi = 0 $ and $ \partial_{\mu}(\phi^{+}\phi) \neq 0 $ then we have the constraint like before:
\begin{eqnarray}\label{constraint currenr current non abelian}
\partial_{\nu}(\phi^{+}\phi) \epsilon^{abc} T^{b} \bar{\psi}\gamma^{\mu}T^{c}\psi = 0
\end{eqnarray}

\subsection{Case 4: Coupling to magnetic moment type interaction}
We begin with an action that has a coupling to magnetic moment type interaction.
\begin{eqnarray}\label{Action: spinor non abelian }
& S=\int{\bar{\psi}(i\gamma^{\mu}\partial_{\mu}-m -  \zeta \bar{\psi}\gamma^{\mu}T^{a}A^{a}_{\mu}\psi)\psi \,d^{4}x}  
\nn & - \frac{1}{4} \int{F^{a\mu\nu}F^{a}_{\mu\nu}\,d^{4}x}
\nn & +\int d^{4}x [ \mu(\phi)D^{A}_{\nu}(\bar{\psi}\sigma^{\mu \nu}\psi)(A_{\mu}+ \frac{i}{\zeta} U\partial_{\mu}U^{-1})
\nn & +\frac{1}{2}\partial_{\mu}\phi\partial^{\mu}\phi - V(\phi)]
\end{eqnarray} 
where $ \sigma^{\mu\nu}=\frac{i}{4}[\gamma^{\mu} , \gamma^{\nu}] $ is the anti - symmetric Dirac tensor and $ D^{A}_{\mu} = \partial_{\mu} + i\zeta T^{a}A^{a}_{\mu} $.
This action is invariant under gauge transformation that also involve the transformation of $U$ 
\begin{equation}\label{Utransf-Cornwalleq 4}
U \rightarrow V U
\end{equation}
so:
\begin{eqnarray}
& \frac{i}{\zeta} U(\theta)\partial_{\mu}U^{-1}(\theta) \rightarrow V(\frac{i}{\zeta} U(\theta)\partial_{\mu}U^{-1}(\theta)) V^{-1} 
\nn & + \frac{i}{\zeta} V \partial_{\mu}V^{-1}
\end{eqnarray}
while $T^{a}A^{a}_{\mu}$ transforms as
\begin{equation}\label{Atransf-Cornwalleq 4}
 T^{a}A^{a}_{\mu} \rightarrow V  T^{a}A^{a}_{\mu}V^{-1} - \frac{i}{\zeta} V \partial_{\mu}V^{-1} 
\end{equation}
The $ F^{a}_{\mu\nu} $ is defined:
\begin{equation}
F^{a}_{\mu\nu}=\partial_{\mu}A^{a}_{\nu} - \partial_{\nu}A^{a}_{\mu} + \zeta f^{a b c}A^{b}_{\mu} A^{c}_{\nu}
\end{equation}
Variation by $ U $ gives the equation:
\begin{eqnarray}\label{case 4 bound U}
& Tr(T^{a}U\delta U^{-1})(\partial_{\mu}\mu(\phi))\bar{\psi}T^{a}\sigma^{\mu\nu}\psi 
\nn & + \mu(\phi)[Tr(T^{a}U\delta U^{-1})\partial_{\mu}(\bar{\psi}T^{a}\sigma^{\mu\nu}\psi)
\nn & + \bar{\psi}T^{a} Tr([T^{a}, \partial_{\mu} U ]\delta U^{-1}])\sigma^{\mu\nu}\psi ] = 0
\end{eqnarray}
We can see that when $ \mu(\phi) = 0 $ and $ \partial_{\mu}\mu(\phi) \neq 0 $ then we have the constraint:
\begin{eqnarray}
(\partial_{\mu}\mu(\phi))\bar{\psi}T^{a}\sigma^{\mu\nu}\psi = 0
\end{eqnarray}
We can see that this constraint is parallel to the constraint of equation \ref{case 1 bound U}.
Someone can check that if we take the action:
\begin{eqnarray}\label{Action: spinor non abelian 2 }
& S=\int \lbrace{\bar{\psi}(i\gamma^{\mu}\partial_{\mu}-m -  \zeta \bar{\psi}\gamma^{\mu}T^{a}A^{a}_{\mu}\psi)\psi}  
\nn & - \frac{1}{4} \int{F^{a\mu\nu}F^{a}_{\mu\nu}\,d^{4}x}
\nn & + (\partial_{\mu}\phi^{+} + i\zeta \phi^{+} T^{a}A^{a}_{\mu})(\partial^{\mu}\phi  - i\zeta T^{a}A^{a\mu}\phi ) - V(\phi^{+}\phi)  
\nn & +  iD^{(\textbf{A})}_{\mu}(\bar{\psi}T^{a}\sigma^{\mu\nu}\psi) \,(\phi^{+} T^{a}\partial_{\mu}\phi - \partial_{\mu}\phi^{+}\,T^{a}\phi 
\nn &- 2\zeta i\epsilon^{abc} \phi^{+} T^{b} B^{c}_{\mu} \phi ) -\frac{1}{4}\,B^{a}_{\mu\nu}B^{a\mu\nu} 
\nn & -\frac{\lambda}{2}\,B^{a}_{\mu\nu}F^{a\mu\nu} \rbrace \,d^{4}x
\end{eqnarray} 
we can get parallel result as equation \ref{constraint currenr current non abelian}
\section{Example of a bag model: The scalar field potetial and the dynamical coupeling constant $ g(\phi) $}\label{sec:The bubble scalar field}
As we saw in case 1 and case 2, we had a situation where $ \partial_{\mu}g(\phi)\bar{\psi}T^{a}\gamma^{\mu}\psi = 0 $ when $ g(\phi)=0 $ and $ \partial_{\mu} g(\phi) = n_{\mu} \neq 0 $. In other words, the scalar field can make domain wall for the Dirac field (the Dirac field cant pass the domain). Lets build bag model by possessing spherical symmetric real static scalar field, which has two vacuum expectation value ($ \phi_{0} $ and $ \phi_{+} $). There is infinitesimal shell on the transition which $ g(\phi_{transition})=0 $ and $ \partial_{\mu} g(\phi_{transition}) \neq 0 $.
Variation on the action \ref{Dirac:boundary20} (case 1) by $ \phi $ gives
\begin{eqnarray}
&\lbrace \frac{\partial\mathcal{L}_{\phi}}{\partial \phi} - \partial_{\mu}\frac{\partial \mathcal{L}_{\phi}}{\partial (\partial_{\mu}\phi)} \rbrace
\nn & +\frac{\partial g(\phi)}{\partial \phi}\bar{\psi}\gamma^{\mu}( T^{a}A^{a}_{\mu} +  \frac{i}{\zeta} U(\theta)\partial_{\mu}U^{-1}(\theta))\psi = 0
\end{eqnarray}
and variation on the action \ref{Dirac:boundary non Abilean} (case 2) by $ \phi $ gives:
\begin{eqnarray}
&\lbrace \frac{\partial\mathcal{L}_{\phi}}{\partial \phi} - \partial_{\mu}\frac{\partial \mathcal{L}_{\phi}}{\partial (\partial_{\mu}\phi)} \rbrace
\nn & +\frac{\partial g(\phi)}{\partial \phi}\bar{\psi}\gamma^{\mu}(\frac{T^{a}A^{a}_{\mu} +  \frac{i}{\zeta} U(\theta)\partial_{\mu}U^{-1}(\theta)}{1+g(\phi)})\psi = 0
\end{eqnarray}
In both cases, outside the bag the term $ \bar{\psi}\gamma^{\mu}( T^{a}A^{a}_{\mu} +  \frac{i}{\zeta} U(\theta)\partial_{\mu}U^{-1}(\theta))\psi = 0 $.
We define the Lagrangian to be:
\begin{eqnarray}
\mathcal{L}_{\phi} = \partial_{\mu}\phi \partial^{\mu}\phi - V(\phi)
\end{eqnarray}
We will follow the nice choice of Friedberg and Lee \citep{FriedbergandLee, FriedbergandLee1} for the potential.
\begin{eqnarray}
V(\phi)= \frac{z \phi_{+}}{2}\phi^{2} - \frac{\phi_{+} + z}{3} \phi^{3} + \frac{\phi^{4}}{4} - \Omega
\end{eqnarray}
where:
\begin{eqnarray}
\Omega = \frac{z \phi_{+}}{2}\phi^{2}_{+} - \frac{\phi_{+} + z}{3} \phi^{3}_{+} + \frac{\phi^{4}_{+}}{4}
\end{eqnarray}
where $ \phi_{0}=0 $ and always $ V(\phi_{+})= 0 $ , and $ \frac{\partial V(\phi_{+})}{\partial \phi} = \frac{\partial V(\phi_{0})}{\partial \phi} = 0 $, where also $ \frac{\partial V(z)}{\partial \phi} =0 $ . Because we want $ \phi_{+} > \phi_{0} $ and $ V(\phi_{+}) < V(\phi_{0}) $ and $ V(\phi_{0}) $ to be a local minimum (this means that $ V(\phi_{0+}) $ is a global minimum), then we need that (see figure 1):
\begin{eqnarray}
0 < z < \frac{\phi_{+}}{2}
\end{eqnarray}

 We take $ g(\phi) $ to be:
\begin{eqnarray}\label{eq:g(phi)}
g(\phi)= \alpha (V(\phi) - C)
\end{eqnarray}
so $ \frac{\partial g(\phi_{0})}{\partial \phi} = \frac{\partial g(\phi_{+})}{\partial \phi} =0 $, and $ \alpha $ is some constant, and $ C $ is a constant which:
\begin{eqnarray}
0<C<\Omega
\end{eqnarray}
We should note that equation \ref{eq:g(phi)} is an example of function that give bag model, but there is also anther kind of functional that can be used.
 Inside the bag, the lowest energy density will be when $ \phi = 0 $, then also the Dirac field don't effect the equation of motion of the scalar field (because $ \frac{\partial g(0)}{\partial \phi}=0 $ ). At the transition from $ \phi_{0} $ to $ \phi_{+} $, there must be a shell where $ g(\phi_{transition})=0 $, so there must be a shell across which the Dirac current has zero flux. Outside the bag the energy density is negative (In comparison to the inside energy density), because the scalar field outside the bag is homogeneous static, and the potential is negative, the true vacuum will pressure to squeeze the bag. At the transition the Dirac field may affect the scalar field, but we can assume that it wont stop the transition of the scalar field.
The size of the bag depends on the amount of pressure of the Dirac field inside the bag, which balance the positive pressure of the scalar field outside the bag. In other word the Dirac field will balance the shrinking of the bag, so it will get a finite size.
\begin{figure}
\includegraphics[scale=.4]{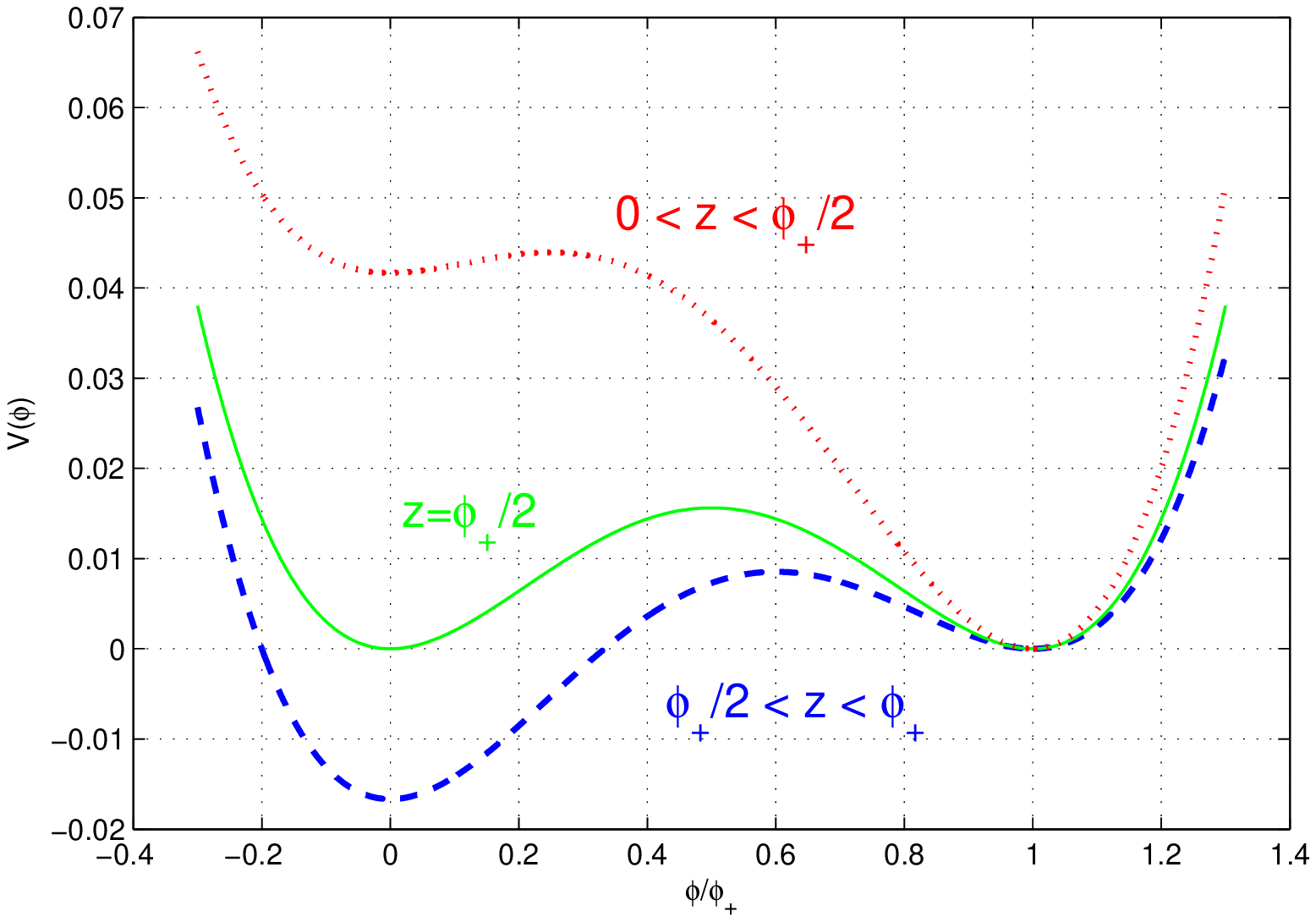} 
\end{figure}
\section{Conclusion}
In this paper we presented a model which can produce confining boundary condition on Dirac field interacting with Abelian or non Abelian gauge fields. The constraint is generated by a scalar field. This kind of model can be the foundation for bag models which can produce confinement. The present work represents among other things a generalization to the non Abelian case of our previous result. In the $U(1)$ case the coupling to the gauge field  contains a term of the form $g(\phi)j_\mu (A^{\mu} +\partial^{\mu}B)$ where $B$ is an auxiliary field and  $j_\mu$ is the Dirac current. The scalar field $\phi$ determines the local value of the coupling of the gauge field to the Dirac particle. The consistency of the equations determines the condition $\partial^{\mu}\phi j_\mu  = 0$
which implies that the Dirac current cannot have a component in the direction of the gradient of the scalar field. As a consequence, if  $\phi$ has a soliton behavior, like defining a bubble that connects two vacuua, we obtain that the Dirac current cannot have a flux through the wall of the bubble, defining a confinement mechanism where the fermions are kept inside those bags. In this paper we presented more models in Abelian case which produce constraint on the Dirac or scalar current and also spin or to the electric and magnetic field. We generalized this procedure for the non Abelian case and we found a constraint that can be used to build a bag model. In the non Abelian case the confining boundary conditions hold at a specific surface of a domain wall.
As we saw in the Abelian case, by building the dielectric field strength in equation \ref{eq:dielectric field strength} or by building the four point interaction \ref{Action:general phase control on Dirac}, we got a constraint on the Dirac current \ref{constraint: Dialectric potetial} and, respectively \ref{constraint:4 point interaction}.
We can also get a constraint on the field strength \ref{constraint: electric field 1} and the Dirac current or respectively the scalar current by using the simple action \ref{action: electric and magnetic field} and respectively \ref{action: electric and magnetic field 2} which give a model that give the M.I.T confinement condition for gauge fields.
Further more, we can have the constraint on the magnetic current \ref{constraint:spinor} and \ref{constraint:spinor 2} by using the action \ref{Action: spinor1 } and, respectively \ref{Action:Dirac spinor 2 }.
We continue with the same procedure also in the non Abelian field and got a parallel result as in the Abelian case.\\
We can generalize this way to get charge confinement (by enforcing that the flux through a surface is vanishes) is to make the coupling constant dynamical. In order to make this consistent with gauge invariance an auxiliary field has to be introduced. The variation of the action with respect to this auxiliary field or the divergence of the gauge field equation produce the confinement condition (a zero flux through the surface condition).
This kind of current constraint can produce different bags models which have different interpretation.\\\\
Finely it is interesting that a connection has been made by other authors between massive gluons and the confinement \citep{Chaichian2006bn, Cornwall}. In our case we see that confinement is associated with dynamical couplings which require the introduction of auxiliary fields, $ B $ or $ U $ depending if we study the abelian or non abelian case.
The same auxiliary field can than be used to construct gauge invariant mass terms, so indeed from this point of view a connection between massive gluons and confinement seems indeed natural.

\bibliography{bagbibeps}

\end{document}